\documentclass{article} %

\usepackage{iclr2026_conference,times}

\iclrfinalcopy

\usepackage{amsmath}
\usepackage{amssymb}
\usepackage{amsfonts}       %
\usepackage{nicefrac}       %

\usepackage{booktabs}       %
\usepackage{threeparttable}
\usepackage{multirow}
\usepackage{longtable}
\usepackage[table,xcdraw]{xcolor}

\usepackage{graphicx}
\graphicspath{{figures/}}   %
\usepackage{float}
\usepackage{placeins}

\usepackage[utf8]{inputenc} %
\usepackage[T1]{fontenc}    %
\usepackage{microtype}      %
\usepackage{soul}           %
\usepackage[normalem]{ulem} %

\usepackage{hyperref}       %
\usepackage{url}            %
\usepackage{cleveref}       %

\title{
uFlowCSP: Crystal Structure Prediction using Mean flow generative models
}

\author{
Sourin Dey\\
 Department of Computer Science and Engineering\\
  University of South Carolina\\
  Columbia, SC 29201 \\
    \And
Dipannoy Das Gupta\\
 Department of Computer Science and Engineering\\
  University of South Carolina\\
  Columbia, SC 29201
    \And
Lai Wei\\
 Department of Computer Science and Cybersecurity\\
  University of North Georgia\\
  Columbia, SC 29201 \\   
  \And
Sadman Sadeed Omee\\
 Department of Computer Science and Engineering\\
  University of South Carolina\\
  Columbia, SC 29201 \\   
  \And 
    Jianjun Hu* \\
 Department of Computer Science and Engineering\\
  University of South Carolina\\
  Columbia, SC 29201 \\
  \texttt{jianjunh@cse.sc.edu} \\
}
\hypersetup{
    colorlinks=true,
    allcolors=blue,
}

\begin{document}
\maketitle
\fancyhead{}

\begin{abstract}
Crystal structure prediction (CSP) is a fundamental task in computational materials
discovery. Generative models such as CDVAE, DiffCSP, FlowMM, and
CrystalFlow have made it possible to learn the distribution of stable crystals
directly instead of searching the potential-energy surface. However, inference remains
the bottleneck in these models: diffusion and flow-matching samplers integrate a
generative trajectory over tens to thousands of sequential network evaluations
per candidate. We introduce uFlowCSP, a deep learning based CSP model built on MeanFlow, which learns
the average rather than the instantaneous velocity of the probability-flow
trajectory and can therefore generate a complete structure in one to five
network evaluations, achieving a $5\times$--$58\times$ speedup with equal or better performance. The velocity field is
parameterized by a chemistry- and symmetry-aware Transformer conditioned on a
canonical atom ordering, a global composition embedding, per-token chemistry
features, and a coarse crystal-system token used only as an auxiliary training
signal while inference remains formula-only. On the MP-20 benchmark, generating 20
candidates per target, a single MeanFlow step matches CrystalFlow's match rate
($78.38\%$ vs.\ $78.34\%$) using $100\times$ fewer network evaluations and
roughly an order of magnitude less wall-clock time; five steps raise this to
$83.64\%$, above CrystalFlow ($78.34\%$ at 2{,}000 evaluations) and DiffCSP
($77.93\%$ at ${\sim}20{,}000$), still at $20\times$ fewer evaluations. uFlowCSP
generates 10k structures in $0.39$--$1.31$\,min, compared with $6.5$\,min for
CrystalFlow and $76.1$\,min for DiffCSP. On the stricter CSPBench protocol, which requires simultaneous structure and
space-group agreement within an energy-ranked top-5 shortlist, and applying
tighter matching tolerances than the MP-20 protocol, uFlowCSP reaches
$72/72/65\%$ structure, space-group, and consensus match
rates at five sampling steps. CrystalFlow attains higher rates ($78/73/68\%$) at
its full 100-step budget, but collapses to $49/32/31\%$ when held to the same
five-step budget, less than half the space-group and consensus rates. uFlowCSP
therefore improves accuracy per network evaluation rather than peak accuracy
alone.
\end{abstract}

\section{Introduction} \vspace{-0.4cm} Crystal structure prediction (CSP) addresses the fundamental challenge of determining how atoms arrange themselves in periodic solids from a given chemical composition alone \citep{oganov2006crystal,oganov2010evolutionary}. The atomic arrangement is the state variable that links composition to behavior: once the structure of a compound is known, first-principles methods can predict its stability, electronic structure, ionic transport, catalytic activity, and mechanical response with quantitative accuracy \citep{oganov2019structure,jain2016computational}. CSP is therefore the enabling step of \emph{in silico} materials discovery, allowing hypothetical compounds, polymorphs, and pressure-stabilized phases to be assessed before they are ever synthesized \citep{needs2016perspective}. Structure searches paved the way to the materials that that lie beyond chemical intuition, encompassing a wide range  from high-temperature superconducting superhydrides to metastable battery cathodes and functional photovoltaic and thermoelectric phases \citep{wang2022crystal,oganov2019structure}. Getting the structure right also means getting its symmetry right, and that symmetry is
specified by the space group and Wyckoff positions. These are physical constraints rather than descriptive labels: they fix site multiplicities and local environments, dictate which properties and electronic band degeneracies are allowed, and underpin the symmetry indicators of electronic topology \citep{wieder2021topological}. A prediction that is geometrically close but crystallographically wrong can imply the wrong physics, so faithful recovery of symmetry is a key requirement for any CSP method \citep{martirossyan2025all}.

Historically, CSP was formulated as a global optimization problem on a potential-energy
surface, where candidate atomic arrangements and lattice parameters were sampled over a
large configurational space and then relaxed and ranked by first-principles energy
calculations \citep{woodley2008crystal,oganov2006crystal,pickard2011ab,wang2010crystal}.
Different search strategies explore this space in different ways: ab initio random structure searching samples configurations stochastically before relaxing them with density functional theory (DFT) \citep{pickard2011airss}, evolutionary algorithms such as USPEX evolve a population toward low-energy minima \citep{oganov2006uspex}, and particle-swarm optimization, as in CALYPSO, navigates the energy landscape using swarm intelligence \citep{wang2010calypso}. These methods are physically rigorous and remain important tools in high-pressure and exploratory materials science, but their reliance on repeated DFT relaxations makes them computationally expensive, a cost that grows sharply with the number of atoms, the unit-cell size, and the size of the chemical search space. 

The advent of deep generative modeling reframed CSP as learning the distribution of stable crystal structures from materials databases. Instead of searching the energy landscape for each new material, the model pays this cost once during training and then generates structures directly. The Crystal Diffusion Variational Autoencoder (CDVAE) pioneered this direction by combining a variational autoencoder with a diffusion-based decoder that denoised crystal representations toward plausible, low-energy configurations \citep{xie2022cdvae}. DiffCSP subsequently formulated CSP as a joint equivariant diffusion process over fractional coordinates and lattice parameters, using periodic E(3)-equivariant modeling to respect the symmetries of crystals \citep{jiao2023diffcsp}. DiffCSP++ incorporated explicit space-group and Wyckoff-position constraints into the generative process, further narrowing the search space to symmetry-compatible structures \citep{jiao2024diffcsppp}.

Crystallographic symmetry has also become an explicit design principle in generative CSP. CrystalFormer uses a space-group-informed autoregressive Transformer that generates symmetry-inequivalent atoms through Wyckoff positions, reducing the effective complexity of crystal generation \citep{cao2024crystalformer}. SymmCD instead decomposes a crystal into its asymmetric unit and the symmetry transformations needed to reconstruct the full structure, allowing diffusion to operate in a compact symmetry-aware representation \citep{levy2025symmcd}. WyckoffDiff uses a discrete diffusion model over space groups and Wyckoff-position occupancies, enforcing symmetry by construction rather than predicting unconstrained atomic coordinates \citep{kelvinius2025wyckoffdiff}. Beyond unconditional or composition-conditioned CSP, MatterGen demonstrated property-conditioned diffusion for steering crystal generation toward target chemistries and properties across the periodic table \citep{zeni2025mattergen}, while autoregressive language models such as CrystaLLM showed that crystals can also be generated directly as text by modeling the Crystallographic Information File (CIF) format \citep{antunes2024crystallm}.

Flow-based generative models provide an alternative to diffusion for learning distributions over periodic materials. FlowMM \citep{miller2024flowmm} introduced Riemannian flow matching for crystal generation, modeling lattice parameters, periodic fractional coordinates, and atom types on manifolds that reflect crystal-specific geometry and symmetries . CrystalFlow \citep{luo2025crystalflow} combined conditional flow matching with a graph-based equivariant network and symmetry-aware crystal representations to generate lattice parameters, atomic coordinates, and compositions, while requiring substantially fewer integration steps than diffusion-based models. These approaches improve sampling efficiency, but they still generally require numerical integration over many network evaluations.

The same limitation holds whether the model is diffusion-based, autoregressive, or flow-based: generation remains a sequential process. Diffusion models denoise over hundreds to thousands of steps, flow-matching models integrate a velocity field over many network evaluations, and autoregressive models emit a structure token by token, so richer symmetry priors and conditioning improve quality without removing the per-candidate inference cost. Reducing that cost to a single or a handful of network evaluations is what motivates one-step generative modeling for CSP. Our uFlowCSP algorithm  characterizes a generative trajectory through an average velocity over a time interval rather than an instantaneous velocity, enabling one-step crystal structure generation without distillation or pre-training \citep{geng2025meanflow}.  %

Evaluating such a model calls for two complementary tests, and we adopt them in sequence. We first place uFlowCSP on the standard MP-20 benchmark \citep{xie2022cdvae,jiao2023diffcsp}, the common proving ground on which diffusion and flow-matching models such as DiffCSP, FlowMM, and CrystalFlow report their any-of-$k$ match rates. Evaluating under this identical protocol allows us compare uFlowCSP head-to-head with those baselines and answer a concrete question: can a one-step or few-step generator match the accuracy of samplers that integrate hundreds to thousands of network evaluations? The match rate alone, however, rewards geometric proximity and can be inflated by duplicate- or polymorph-contaminated splits, so a high score is not the only factor that indicates a model has learned the underlying crystallography \citep{martirossyan2025all}. We therefore turn to CSPBench \citep{wei2024cspbench}, a curated suite of 180 held-out structures that span binary to quaternary compositions and graded difficulty and are scored not only by structure matching but also by space-group match rate and their consensus match rates. Because CSPBench withholds symmetry information at inference and credits a prediction only when structure and space group agree simultaneously, it directly probes whether a model recovers the correct symmetry, which is what our symmetry-aware conditioning targets. This two-stage evaluation (efficiency and accuracy on MP-20, symmetry fidelity on CSPBench) frames the results reported below.

This work makes two contributions. First, we adapt one-step MeanFlow generation to crystal structure prediction, showing that a complete periodic structure can be produced in a single network evaluation rather than by integrating a generative trajectory. Our method achieves a higher match rate than multi-step diffusion and flow-matching baselines at 20 to 200 times fewer network function evaluations and 5 to 58 times lower measured wall-clock time. Second, we introduce a chemistry- and symmetry-aware Transformer parameterization of the average-velocity field, and show by ablation that canonical atom ordering, a global composition embedding, per-token chemistry features, and a coarse crystal-system auxiliary token contribute additively. Especially, we find that the auxiliary symmetry token introduced only during training stage can  improve the space-group agreement despite it remaining unused at inference.

\begin{figure*}[t]
  \centering
  \includegraphics[width=\textwidth]{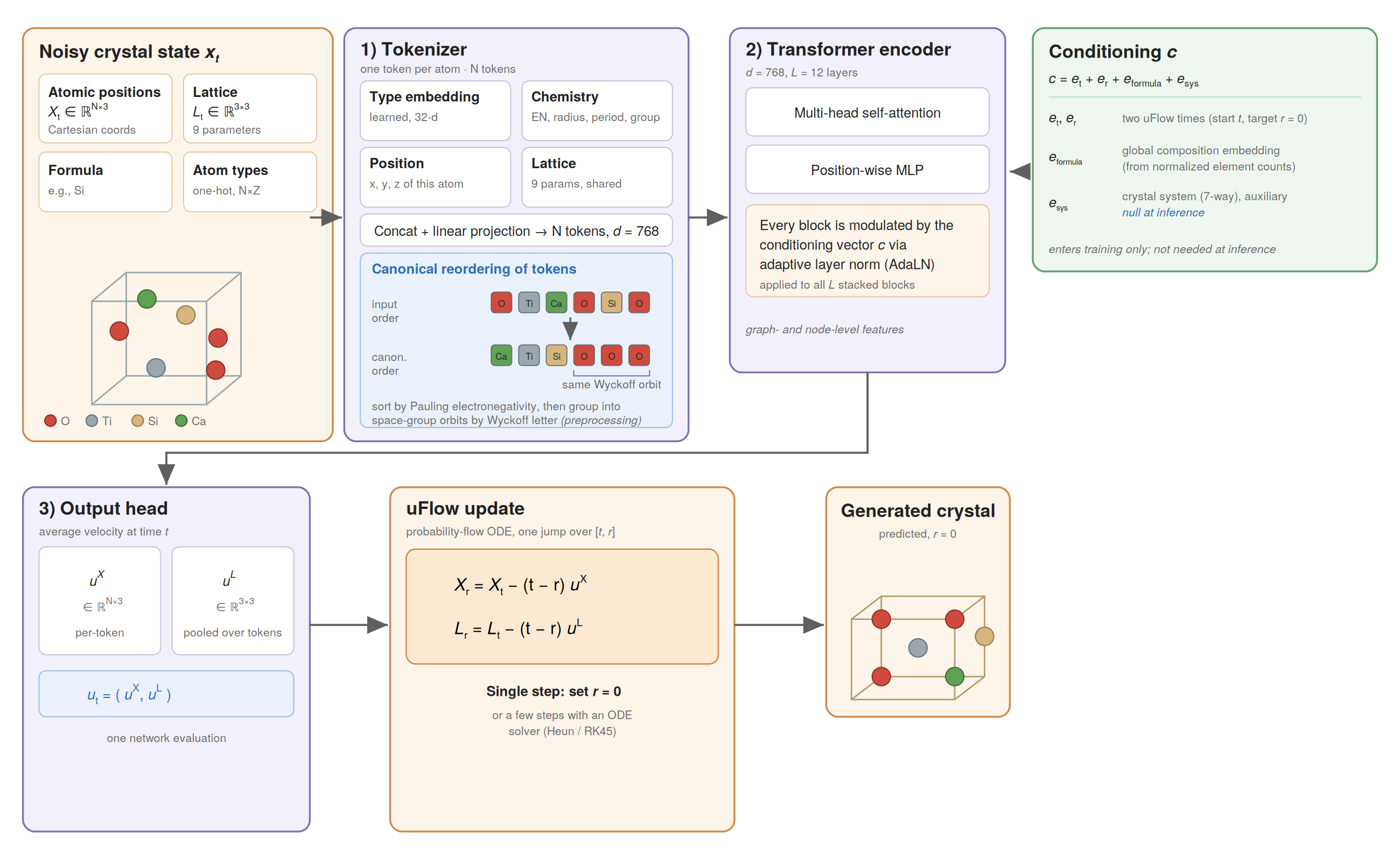}
\caption{\textbf{uFlowCSP architecture.} A crystal is encoded as one token per atom, each carrying the atom's Cartesian position and a shared copy of the flattened lattice, together with a learned type embedding and four fixed chemistry descriptors (electronegativity, radius, period, group). Tokens are sorted in a canonical order by electronegativity, and then grouped into Wyckoff orbits. This input is processed by a Transformer whose blocks are modulated through adaptive layer normalization by a global conditioning vector $\mathbf{c}=\mathbf{e}_t+\mathbf{e}_r+\mathbf{e}_{\mathrm{formula}}+\mathbf{e}_{\mathrm{sys}}$, formed from the two MeanFlow time embeddings, a composition embedding, and a crystal-system token used only during training. The output head predicts the interval-\emph{averaged} velocity, whose coordinate and lattice components define a single MeanFlow update that carries a noisy state directly to a generated crystal in one step (or a few). At inference every conditioning signal is derived from composition alone; the space group is never supplied.}
  \label{fig:architecture}
\end{figure*}

\section{Method}
\label{sec:method}

\subsection{Problem Setting and Representation}
We consider crystal structure prediction under two evaluation protocols (MP20 and CSPBench), both of which supply the primitive-cell atom multiset as the composition condition, so the number of atoms is known in either case, and neither supplies a space-group label at inference. A crystal containing $N$ atoms is represented in raw, all-atom form by its atomic numbers $\{z_i\}_{i=1}^{N}$, Cartesian coordinates $\mathbf{X}\in\mathbb{R}^{N\times 3}$, and lattice matrix $\mathbf{L}\in\mathbb{R}^{3\times 3}$. The lattice matrix is flattened into nine continuous parameters. Atomic numbers are fixed by the composition condition, while the model predicts the continuous coordinates and lattice parameters.

\subsection{Few-Step Generative Backbone}
\label{sec:backbone}

We generate crystals with a MeanFlow model \citep{geng2025meanflow}, a flow-based generator that produces crystal structures in a small number of network evaluations. We first describe how a crystal is encoded as the state the flow acts on, briefly recall the MeanFlow formulation, and then give the crystal-specific velocity prediction and sampling procedure.

\subsubsection{Crystal Representation}

We represent a crystal of $N$ atoms as a set of $N$ tokens. Each token concatenates the atom's position with the flattened unit cell,
\[
\mathbf{x}_i = \big[\, \mathbf{r}_i \;;\; \operatorname{vec}(\mathbf{L}) \,\big] \in \mathbb{R}^{12},
\qquad i = 1, \dots, N,
\]
where $\mathbf{r}_i \in \mathbb{R}^{3}$ are Cartesian coordinates (in \AA) and $\operatorname{vec}(\mathbf{L}) \in \mathbb{R}^{9}$ is the row-major flattening of the lattice matrix $\mathbf{L} \in \mathbb{R}^{3\times3}$. Stacking the tokens gives the state $\mathbf{x} \in \mathbb{R}^{N \times 12}$ on which the flow operates. Atomic numbers $\{z_i\}$ are fixed by the composition and enter the network as conditioning rather than being generated (Section~\ref{sec:transformer}); the flow transports only the continuous coordinates and lattice.

Every token carries a copy of the same lattice. This redundancy lets the Transformer read the cell at every position through self-attention without a separate lattice pathway, at the cost of replicating the nine lattice channels $N$ times in the state. Because all copies describe one physical cell, we reduce them to a single lattice at the end of sampling by averaging over atoms; the full recovery procedure is given in the Supplementary Materials.

\subsubsection{MeanFlow}

MeanFlow \citep{geng2025meanflow} is a flow-based generator built to sample in few network evaluations; we summarize only what is needed below and refer to \citep{geng2025meanflow} for the derivations. It places data $\mathbf{x}_1$ and noise $\boldsymbol{\epsilon}\sim\mathcal{N}(0,\mathbf{I})$ at the two ends of a linear path and defines the state at time $t$ as their interpolation,
\[
\mathbf{z}_t = (1-t)\,\mathbf{x}_1 + t\,\boldsymbol{\epsilon},
\qquad t \in [0,1],
\]
so $\mathbf{z}_t$ is a data--noise mixture whose endpoints are pure noise at $t=1$ and pure data at $t=0$, with instantaneous velocity $\mathbf{v} = \tfrac{d\mathbf{z}_t}{dt} = \boldsymbol{\epsilon} - \mathbf{x}_1$. Rather than the instantaneous velocity of ordinary flow matching, which must be integrated over many steps at sampling time, MeanFlow learns the \emph{average} velocity $\mathbf{u}(\mathbf{z}_t, r, t) = \tfrac{1}{t-r}\int_r^t \mathbf{v}\,d\tau$ over an interval, so that any two times are related in one step,
\[
\mathbf{z}_r = \mathbf{z}_t - (t-r)\,\mathbf{u}(\mathbf{z}_t, r, t).
\]
The network $\mathbf{u}_\theta$ is trained to satisfy the MeanFlow identity (Eq.~6 of \citep{geng2025meanflow}), $\mathbf{u} = \mathbf{v} - (t-r)\tfrac{d}{dt}\mathbf{u}$, by regressing onto the stop-gradient target $\mathbf{u}_{\text{tgt}} = \mathbf{v} - (t-r)\tfrac{d}{dt}\mathbf{u}_\theta$ under a squared, adaptively weighted loss; the total derivative $\tfrac{d}{dt}\mathbf{u}_\theta$ is obtained as a Jacobian--vector product via forward-mode automatic differentiation in the same forward pass. 

\subsubsection{Velocity Prediction for Crystals}

In our setting, the network $\mathbf{u}_\theta(\mathbf{z}, r, t)$ takes the crystal state $\mathbf{z} \in \mathbb{R}^{N\times 12}$ and outputs a tensor of the same shape: a per-atom displacement whose first three channels move the atom's coordinates and whose remaining nine channels move that atom's copy of the lattice. It is influenced by the chemistry and symmetry-aware Transformer of Section~\ref{sec:transformer}, conditioned on the composition and on the two times $t$ and $r$, so the predicted mean velocity is informed jointly by all atoms and by the shared cell. Coordinates are transported in Cartesian space along the linear path. During training the crystal-system conditioning is dropped for a fraction of examples, so the same network represents both symmetry-conditioned and unconditioned velocity fields; Section~\ref{sec:transformer} details this mechanism and its interpretation, and sampling uses the network without any guidance.

\subsubsection{Sampling}

To generate a crystal we fix the composition, draw an initial noise state
$\mathbf{z}^{(0)} \sim \mathcal{N}(0,\mathbf{I}) \in \mathbb{R}^{N\times 12}$,
and integrate from $t=1$ (noise) to $t=0$ (data) in $S$ uniform steps.
We use a time schedule $t_0 = 1 > t_1 > \dots > t_S = 0$ with spacing
$\Delta t = 1/S$, and write $\mathbf{z}^{(k)}$ for the state after $k$ steps.
Each step applies the mean-velocity jump
\[
\mathbf{z}^{(k+1)} \;=\; \mathbf{z}^{(k)}
\;-\; \Delta t\;\mathbf{u}_\theta\!\big(\mathbf{z}^{(k)},\, t_{k+1},\, t_k\big),
\qquad k = 0,\dots,S-1,
\]
where the superscript $(k)$ indexes the step, $t_k$ is the current time and
$t_{k+1}$ the target time of the jump (the $(t,r)$ pair of the average velocity
$\mathbf{u}_\theta$), and $\theta$ denotes the trained network weights.
Setting $S=1$ gives single-step generation, the defining capability of MeanFlow;
we report $S=5$ as our main configuration, which improves quality at a still-small,
fixed cost. The final state $\mathbf{z}^{(S)}$ is decoded into a crystal by averaging
its per-atom lattice channels into one cell and reading off the atomic coordinates,
as detailed in the Supplementary Materials.

\subsection{Chemistry- and Symmetry-Aware Transformer}
\label{sec:transformer}
The velocity field is parameterized by a sequence Transformer that treats each
atom as a token. A plain Transformer over raw coordinates and a lattice must
infer all chemistry from the atom-type indices alone, which is a hard starting
point: the same element behaves very differently across compositions, and the
token order carries no meaning. We address this in two ways. First, we fix a
canonical atom order so the order-sensitive attention sees a consistent
sequence. Second, we add three conditioning signals, each drawn only from the
composition so that inference stays formula-only: per-atom chemistry, applied to
each token individually; whole-formula chemistry; and a coarse symmetry hint,
both applied globally. Each is described below, followed by how it enters the
network.

\paragraph{Canonical atom ordering}
Self-attention has no inherent notion of token order, so an arbitrary atom ordering forces the model to treat every permutation of the same crystal as a different input. Following MCFlow \citep{seong2026mcflow}, we remove this nuisance by fixing a canonical order: atoms are sorted by Pauling electronegativity, and for symmetry-aware training examples they are further grouped into space-group orbits and ordered by Wyckoff letter. This ordering is computed once during preprocessing with \textsc{pymatgen} \citep{ong2013pymatgen} and its \textsc{spglib} backend \citep{togo2024spglib}. The symmetry grouping uses the space group and is therefore available only in training; at formula-only inference the atoms are ordered by electronegativity alone, with no space-group label required.

\paragraph{Per-atom chemistry}
The per-atom signal attaches chemical information to each token individually. Each atom already carries a learned atom-type embedding: a trainable vector, looked up by atomic number, that the network optimizes during training as its
internal representation of the element. On its own this embedding must recover all periodic-table structure from data. We therefore concatenate onto each token four fixed descriptors of its element, Pauling electronegativity, atomic radius,
period, and group. These give the network direct access to periodic-table trends rather than forcing it to learn them from atom-type indices. Being per-atom, they act locally, alongside the atom's current coordinates and the shared lattice.

\paragraph{Whole-formula chemistry}
The second signal describes the composition as a whole. We form a vector of element fractions for the crystal and pass it through a small multilayer perceptron to obtain a single embedding $\mathbf{e}_{\mathrm{formula}}$. Where the per-atom features tell the model what each atom is, this tells it what the overall stoichiometry is. For example, that a candidate is an oxide in a $1{:}2$ ratio , which constrains the kind of structure that is plausible. Because it summarizes the entire formula rather than one atom, it is applied globally to every token rather than attached to any one of them.

\paragraph{Coarse symmetry}
The third signal is a coarse hint about symmetry. Rather than the exact space group, which would be a 230-way choice and is not knowable from the formula, we map the space group to just its crystal system: one of seven classes (triclinic, monoclinic, orthorhombic, tetragonal, trigonal, hexagonal, cubic), plus a null class used when no symmetry is supplied. Each class is shared by many space groups and so is seen far more often in training, giving a broad symmetry prior without forcing the model to commit to an exact group. Like the formula embedding, this is a property of the whole crystal and is applied globally as $\mathbf{e}_{\mathrm{sys}}$.

\paragraph{How the signals enter.}
The three signals act at two different levels. The per-atom chemistry features are local: they are concatenated onto each atom's token alongside its learned atom-type embedding $\mathbf{a}_i$, coordinates $\mathbf{r}_i$, and the shared lattice, forming the input token
\[
\mathbf{h}_i
=
\big[\,
\mathbf{a}_i \;;\;
\mathbf{f}_i \;;\;
\mathbf{r}_i \;;\;
\operatorname{vec}(\mathbf{L})
\,\big],
\]
where $\mathbf{f}_i$ are the four fixed descriptors of atom $i$. The two whole-crystal signals are global: together with the two MeanFlow time embeddings they form a single conditioning vector that modulates every Transformer block through adaptive layer normalization,
\[
\mathbf{c}
=
\mathbf{e}_t
+
\mathbf{e}_r
+
\mathbf{e}_{\mathrm{formula}}
+
\mathbf{e}_{\mathrm{sys}},
\]
where $\mathbf{e}_t$ and $\mathbf{e}_r$ embed the two MeanFlow times. In short, atom-wise chemistry enters through the tokens $\mathbf{h}_i$, while formula-wise and system-wise conditioning enter through $\mathbf{c}$.

The crystal-system token is a training aid, not an inference input. Following classifier-free guidance, it is set to its null class for a fraction of training examples, so the network learns to predict velocities both with and without it. At inference the token is always null. The improvement it gives is thus a regularization effect: the symmetry-aware task during training yields a better representation for the formula-only generator. At inference the model uses only atom order, per-atom chemistry, and the formula embedding, all derived from the composition; the space group is not needed to generate a structure.

\subsection{Implementation and Training}
We train on the MP-20 subset of the Materials Project \citep{jain2013materialsproject,xie2022cdvae}, which contains structures with at most 20 atoms per unit cell, using the standard CDVAE split \citep{xie2022cdvae}. The 180 CSPBench evaluation structures are removed from the training set, so no evaluation structure is seen during training. The denoiser is a 12-layer Transformer with hidden size 768 and 12 attention heads; atom types use a 32-dimensional learned embedding. Models are implemented in PyTorch \citep{paszke2019pytorch} and trained for 700 epochs in mixed \texttt{bf16} precision with AdamW at a constant learning rate of $10^{-4}$ and batch size 256 on a single NVIDIA A100 GPU. The MeanFlow flow ratio is $0.25$, the classifier-free-guidance dropout probability is $0.8$, and the guidance scale is $1.5$. The Jacobian--vector products in the MeanFlow objective are computed with forward-mode automatic differentiation.

\paragraph{Training-data scope.}
We evaluate under two protocols: the standard MP-20 benchmark
\citep{xie2022cdvae,jiao2023diffcsp}, which supplies the primitive-cell composition
and scores raw candidates by Match Rate and RMSE (Protocol~B), and CSPBench
\citep{wei2024cspbench}, which supplies the same primitive-cell composition and scores an energy-ranked
top-5 shortlist by structure, space-group, and consensus rates (Protocol~A). We
train a separate model for each, in both cases on MP-20 with the evaluation
structures held out: the 180 CSPBench targets for Protocol~A, and the standard
test split for Protocol~B. We keep training to MP-20 for three reasons.

First, all generative baselines we compare against (CDVAE~\citep{xie2022cdvae},
DiffCSP~\citep{jiao2023diffcsp}, FlowMM~\citep{miller2024flowmm},
CrystalFlow~\citep{luo2025crystalflow}) are trained on the same MP-20 split.
Fixing the training data lets us attribute our gains to the one-step objective
rather than to more data, and keeps our numbers comparable to published results.

Second, the 180 CSPBench targets \citep{wei2024cspbench} are not a random MP-20 test
sample but a curated, difficulty-stratified set consisting of 60 binary, 60 ternary, and 60
quaternary compounds, graded from easy to hard by symmetry, prototype, element count,
and cell size and they are scored on structure match, space-group match, and their consensus rate.

Third, more data does not help a generative model the way it helps a template
method such as TCSP~2.0~\citep{wei2025tcsp}. A template library is a lookup index, so wider coverage always helps. A
generative model instead learns a distribution, and the extra sources are far from
real bulk crystals: GNoME \citep{merchant2023scaling} is mostly DFT-hypothetical,
with only 736 of 2.2 million entries experimentally confirmed, and C2DB
\citep{gjerding2021c2db} is two-dimensional. We leave training on larger, carefully
cleaned data to future work.

\subsection{Evaluation Protocols}
\label{sec:evaluation}
We report results under two protocols that test different things: Protocol~A scores relaxed, energy-ranked candidates and credits a prediction only when both structure and space group are correct, while Protocol~B scores raw generated candidates on structure match alone. Their absolute rates are therefore not directly comparable and are reported in separate tables.

\paragraph{Protocol A: the standard MP-20 evaluation.}\label{protocolB}
This protocol follows the raw, any-of-$k$ MP-20 evaluation adopted by prior generative CSP models, including CDVAE~\citep{xie2022cdvae}, DiffCSP~\citep{jiao2023diffcsp}, and CrystalFlow~\citep{luo2025crystalflow}. The model receives the full unit-cell composition, namely the atom multiset of the primitive cell, so the number of atoms is known. For each structure in the 9,046-structure MP-20 test split, we generate $k$ candidates without relaxation or energy-based selection. Each reference structure is scored only against its own $k$ generated candidates.

A structure is counted as matched if \texttt{StructureMatcher.get\_rms\_dist} returns a value within $\mathrm{ltol}=0.3$, $\mathrm{stol}=0.5$, and $\mathrm{angle\_tol}=10^\circ$ for at least one candidate. The match rate is the fraction of the 9,046 test structures with at least one such match. RMSE is computed from the minimum normalized root-mean-square displacement among the candidates for each matched structure and then averaged over matched structures.

\paragraph{Protocol B: CSPBench \citep{wei2024cspbench}.} \label{protocolA}
CSPBench contains 180 held-out structures; we evaluate them with the energy-ranked top-5 shortlist procedure used by TCSP~2.0 \citep{wei2025tcsp}. Following this protocol, the model receives the primitive-cell chemical formula, fixing the atom multiset and hence the number of formula units $Z$ in the cell, but no space-group label is provided. This is the same composition format used to benchmark TCSP~2.0, CSPML, and EquiCSP in Figure~\ref{fig:tcsp-comparison}, so that comparison is made under matched input information. For each formula, we generate 50 candidate structures using the five-evaluation sampling configuration used for this protocol. Each candidate is locally relaxed with ORB-v3 \citep{rhodes2025orbv3}, a universal machine-learned interatomic potential that ranks among the top performers for structure relaxation and stability prediction on the Matbench Discovery leaderboard \citep{riebesell2025matbench}. The relaxed candidates are ranked by CHGNet-predicted energy per atom \citep{deng2023chgnet}, and the five lowest-energy candidates form the top-5 shortlist. The reference structure is used only for scoring and never for candidate selection, so this is a genuine top-5 metric rather than recall over all 50 generated candidates.

A candidate is counted as a structural match if \texttt{pymatgen}'s \texttt{StructureMatcher} \citep{ong2013pymatgen} accepts it against the reference using $\mathrm{ltol}=0.2$, $\mathrm{stol}=0.3$, and $\mathrm{angle\_tol}=5^\circ$ on primitive cells. Both candidate and reference structures are lightly symmetry-refined at $\mathrm{symprec}=0.1\,\text{\AA}$ before matching, so the structural test is independent of the space-group assignment. A candidate is counted as a space-group match when the integer label from $1$ to $230$ returned by \texttt{SpacegroupAnalyzer} with its \texttt{spglib} backend \citep{togo2024spglib}, evaluated at the same tolerance, equals the label assigned to the reference. The same procedure is applied to generated and reference structures, making the assignment deterministic and symmetric.

The structure-match rate and space-group-match rate are the fractions of the 180 formulas for which at least one top-5 candidate matches the reference under the corresponding criterion. The consensus rate is the fraction of formulas for which one top-5 candidate matches in both structure and space group simultaneously. Formulas for which no valid candidate is produced are counted as failures.

\section{Results \& Discussion}
\label{sec:Results}

\begin{figure}[htbp]
  \centering
  \includegraphics[width=\linewidth]{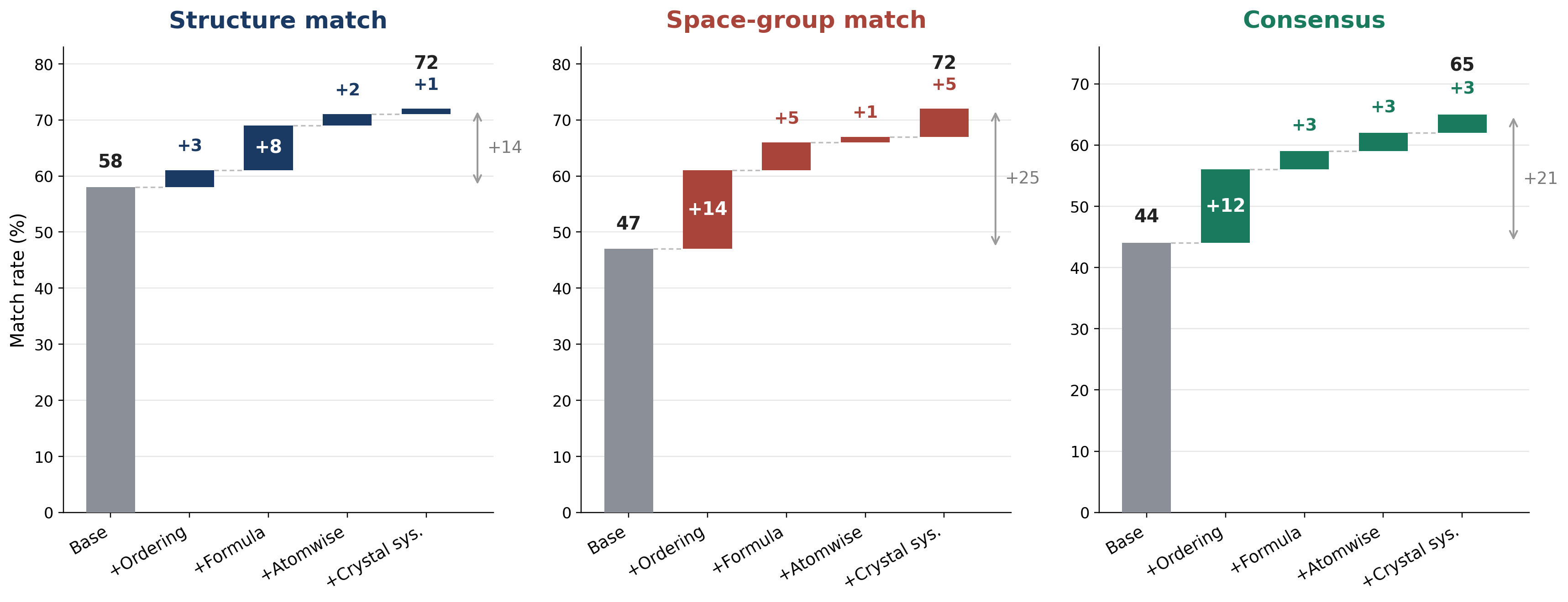}
 \caption{%
  Additive effect of the four conditioning components on uFlowCSP crystal-structure
  prediction. Each panel is
  a separate metric; the grey bar is the baseline (no chemical or symmetry
  priors) and each colored block is the increment contributed by switching on one
  component, in the order they are added, completing the model at $72/72/65$. All rows
  use an identical five-step sampling budget, so the gains reflect conditioning rather
  than extra integration steps.%
}
  \label{fig:ablation-waterfall}
\end{figure}

\begin{figure}[htbp]
  \centering
  \includegraphics[width=\linewidth]{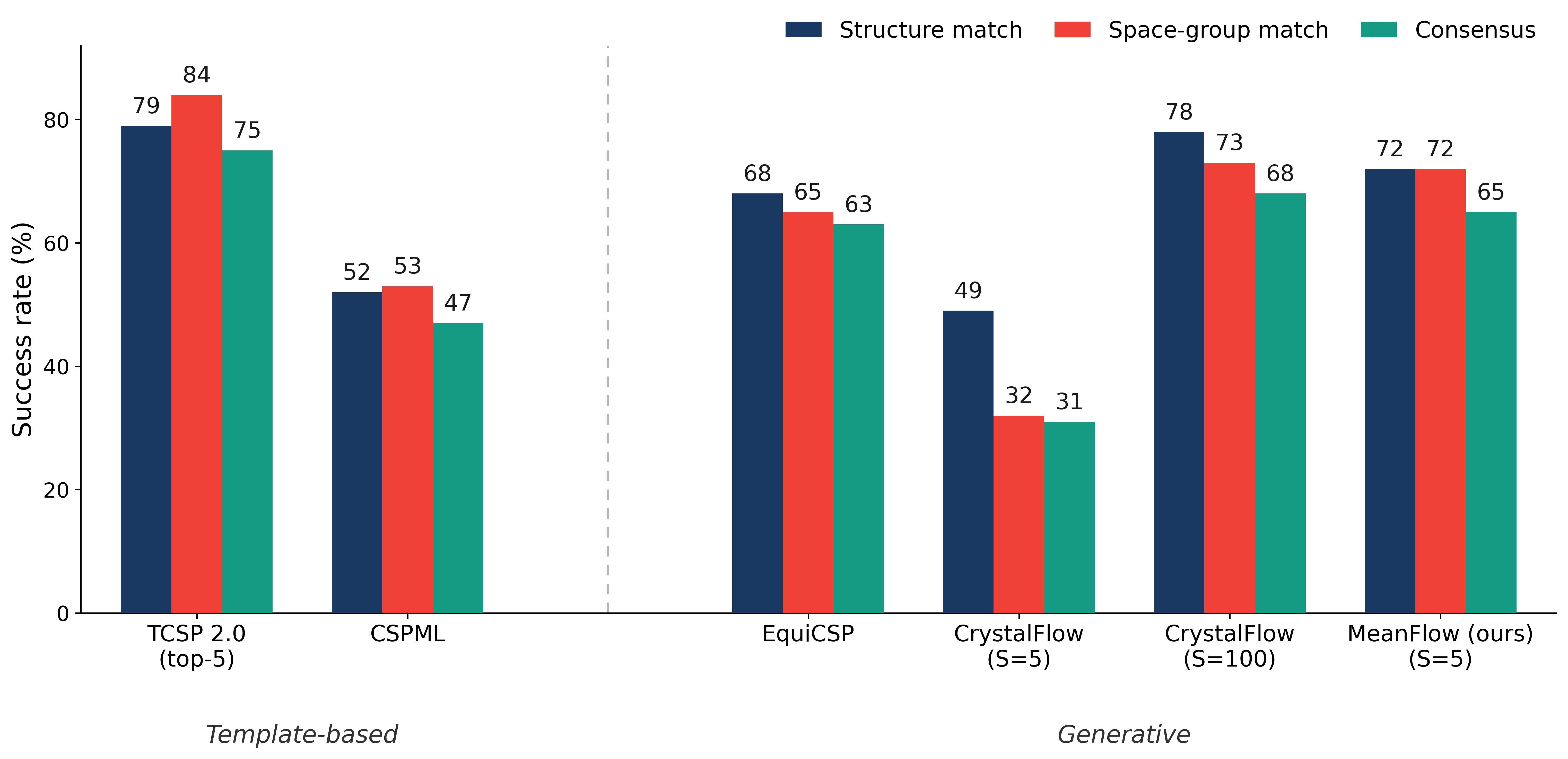}
 \caption{%
    Success rates for structure match, space-group match, and consensus on the
    180-formula CSPBench test set, grouped by method family, under the protocol of
    \Cref{protocolA} (primitive-cell-formula conditioning, space group withheld,
    top-5 selection). A prediction counts toward consensus only when structure and
    space-group match both succeed. Template-based methods (TCSP~2.0, CSPML) and
    generative methods are not matched on data access: TCSP~2.0 substitutes from a
    731{,}293-entry library spanning the Materials Project, Materials Cloud, C2DB, and
    GNoME~\citep{wei2025tcsp}, whereas the generative models are trained on MP-20 alone.
    CrystalFlow is shown at both $S=5$ and its full $S=100$ budget. Values for TCSP~2.0,
    CSPML, and EquiCSP are as published~\citep{wei2025tcsp}; CrystalFlow and uFlowCSP
    were measured here.%
}
  \label{fig:tcsp-comparison}
\end{figure}

\begin{figure}[htbp]
    \centering
    \includegraphics[width=\linewidth]{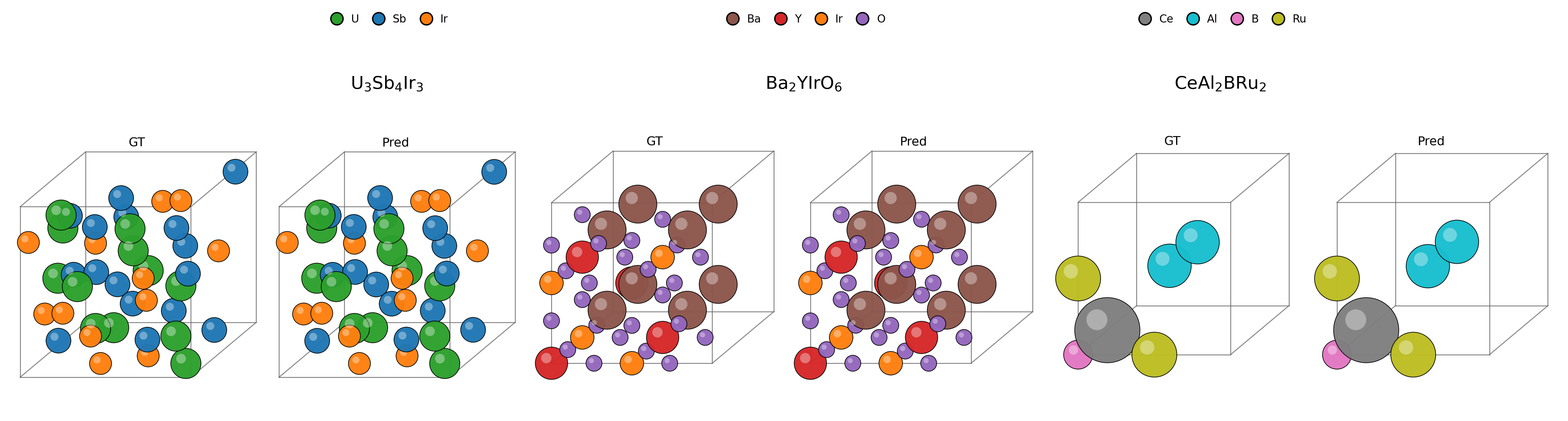}
    \caption{Ground truth (GT) versus predicted (Pred) crystal structures
    for three multi-element consensus examples: (a) U$_3$Sb$_4$Ir$_3$,
    (b) Ba$_2$YIrO$_6$, and (c) CeAl$_2$BRu$_2$. All three achieve both
    StructureMatcher and space-group agreement with the ground truth.
    Predicted structures are shown in the ground truth's coordinate frame
    (translation/axis alignment only, via pymatgen's
    \texttt{StructureMatcher.get\_s2\_like\_s1}, the same transformation
    used internally to establish the match) for direct visual comparison;
    reported match statistics are computed on the unaligned structures.}
    \label{fig:complex-consensus}
\end{figure}

\begin{figure}[htbp]
  \centering
  \includegraphics[width=\linewidth]{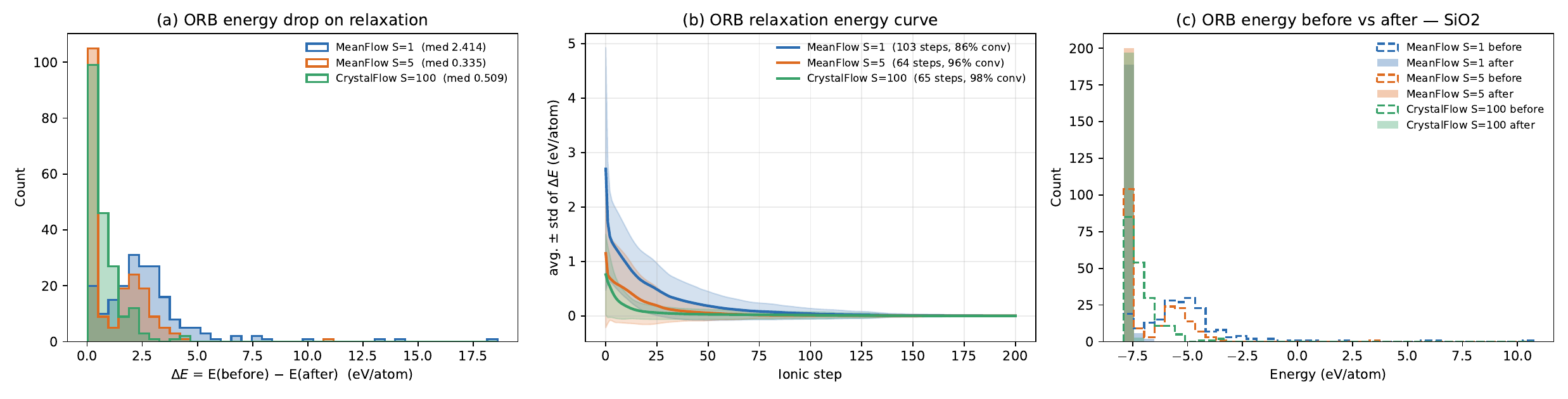}
  \caption{%
    \textbf{ORB-v3 relaxation behavior for SiO$_2$ structures (machine-learned-potential
    proxy).} Structures generated by one-step uFlowCSP ($S{=}1$), five-step uFlowCSP
    ($S{=}5$), and CrystalFlow ($S{=}100$) are relaxed under an identical ORB-v3
    protocol~\citep{rhodes2025orbv3}. All quantities are defined by the potential and
    optimizer, \emph{not} by DFT, and are not directly comparable to the DFT/VASP
    values in the original CrystalFlow study.
    \textbf{(a)} Per-structure energy drop on relaxation
    $\Delta E = E(\text{before}) - E(\text{after})$; legend reports the median
    $\Delta E$. Smaller values indicate raw outputs already near a relaxed local
    minimum. \textbf{(b)} Average ($\pm$ std) $\Delta E$ versus ionic step during
    relaxation; legend reports the mean ionic-step count and ORB-defined convergence
    rate. \textbf{(c)} Energy distributions before (dashed) and after (filled)
    relaxation; the three generators converge to the same low-energy peak. Five-step
    uFlowCSP is competitive with CrystalFlow $S{=}100$ on this proxy at $20\times$
    fewer network evaluations.%
  }
  \label{fig:orb-relax}
\end{figure}

We organize the evaluation around three questions, all under the formula-only
CSPBench protocol except where noted. \emph{(i) What accuracy does the model reach,
and where does it come from?} We isolate the contribution of each conditioning
signal (the ablation of Figure~\ref{fig:ablation-waterfall}), compare the velocity
predictor against an equivariant-GNN backbone (Table~\ref{tab:models}), and place
the full model against template-based and generative baselines, including CrystalFlow
at matched and full cost (Figure~\ref{fig:tcsp-comparison}, Table~\ref{tab:topk_model_comparison});
Figure~\ref{fig:complex-consensus} illustrates the outcome on three individual
structures. \emph{(ii) How does accuracy trade against inference cost?} We place the
model on the accuracy--cost frontier of the standard MP-20 benchmark
(Table~\ref{tab:mp20-main}), where its one- to few-step sampling is directly
comparable to multi-step baselines. \emph{(iii) How close are the raw outputs to a
relaxed minimum?} We report an ORB-v3 relaxation diagnostic on SiO$_2$
(Figure~\ref{fig:orb-relax}). Throughout, the recurring theme is accuracy
\emph{per network evaluation} rather than peak accuracy alone.

\paragraph{Component ablation.}
Figure~\ref{fig:ablation-waterfall} builds the model up one signal at a time, and the
resulting waterfall tells a clear story: the four signals split into two groups, one
that mainly fixes symmetry and one that mainly fixes geometry, and the single largest
gain comes essentially for free. We start from a raw all-atom MeanFlow with no chemical
or symmetry priors, which reaches only $58/47/44$ for structure, space-group, and
consensus match -- it often lands near the right geometry but rarely with the right
symmetry. Adding the canonical atom ordering gives the largest jump anywhere in the
ablation ($+14$ space group, $+12$ consensus), and it is purely a preprocessing choice:
by presenting the order-sensitive Transformer with a deterministic, symmetry-grouped
token sequence, we remove a nuisance degree of freedom the model was otherwise wasting
capacity to absorb. This is the clearest single lesson of the ablation -- most of the
symmetry problem was token order, not model capacity. The chemistry signals then
address the complementary axis: the global formula embedding gives the largest
structure-match gain ($+8$), and the per-token chemistry features add a smaller further
gain, together sharpening geometry rather than symmetry. The coarse crystal-system
token closes the model at $72/72/65$, contributing a final $+5$ to space group; since
it is dropped to a null token at inference (Section~\ref{sec:transformer}), this gain
reflects regularization from the auxiliary symmetry task rather than any symmetry
information used at generation. Every row uses the same five-step budget, so these are
gains from conditioning, not from extra integration steps.

\begin{table}[htbp]
\centering
\caption{Average-velocity predictor: chemistry- and symmetry-aware Transformer
versus an equivariant-GNN backbone, at matched cost. Both are trained under the
identical MeanFlow objective and evaluated with the same $S=5$ integration budget
and top-5, CHGNet-ranker selection protocol; the Transformer uses the best
configuration from the ablation in Figure~\ref{fig:ablation-waterfall} (atom
\textbf{Ord}ering + \textbf{Form}ula embedding + per-token \textbf{Atom}wise
chemistry features + coarse crystal-\textbf{sys}tem auxiliary conditioning).
Results are percentages; best in \textbf{bold}. CrystalFlow at matched ($S=5$)
and full ($S=100$) budgets is reported separately in
Table~\ref{tab:topk_model_comparison}.}
\label{tab:models}
\begin{tabular}{lcc}
\toprule
 & \multicolumn{2}{c}{\textbf{Average-velocity predictor}} \\
\cmidrule(lr){2-3}
\textbf{Metric} & \textbf{Transformer} & \textbf{Equivariant GNN} \\
                & \textbf{(Ord+Form+Atom+Sys)} & \\
\midrule
Structure match rate   & \textbf{72} & 70 \\
Space-group match rate & \textbf{72} & 64 \\
Consensus rate         & \textbf{65} & 60 \\
\bottomrule
\end{tabular}
\end{table}

\paragraph{Velocity-field parameterization.}
The ablation varied the conditioning signals while keeping the predictor fixed, which
isolates the effect of conditioning. To isolate the effect of the predictor itself,
we now keep the conditioning fixed at its best setting and vary the backbone.
Table~\ref{tab:models} reports this comparison. In place of the Transformer, we
evaluate an equivariant graph neural network of the DiffCSP family
\citep{jiao2023diffcsp}, trained with the same MeanFlow objective and the same
$S=5$ budget, so that the predictor is the only difference. The Transformer performs
better ($72/72/65$ versus $70/64/60$), most clearly on symmetry.

The more demanding comparison is against CrystalFlow, a strong flow-matching
baseline, at equal cost. Given the same five steps, CrystalFlow reaches only
$49/32/31$ (Table~\ref{tab:topk_model_comparison}), less than half the space-group
and consensus rates of uFlowCSP. Flow matching degrades substantially when the
trajectory is integrated in only a few steps, whereas the average-velocity
formulation does not; this is precisely where MeanFlow is advantageous, as it
preserves symmetry at low step counts. CrystalFlow surpasses uFlowCSP only when its
budget is increased twenty-fold to $S=100$ ($78/73/68$), and even then its advantage
on symmetry is marginal ($73$ versus $72$) at twenty times the number of network
evaluations. In summary, uFlowCSP trades a small reduction in maximum attainable
accuracy for a substantial gain in inference efficiency.

\begin{table}[htbp]
\centering
\caption{Top-$k$ performance of uFlowCSP and CrystalFlow on the 180-formula CSPBench
split following Protocol~A (see \ref{protocolA}). uFlowCSP uses $S=5$ integration steps. CrystalFlow is designed for many-step
integration, so we report it both at its full $S=100$ budget and at a matched $S=5$
budget for reference; the $S=5$ column reflects operation well below its intended
step count rather than a like-for-like weakness, and it recovers its strong accuracy
at $S=100$, using $20\times$ more network evaluations.}
\label{tab:topk_model_comparison}
\small
\begin{tabular}{lccccccccc}
\toprule
\textbf{Model} 
& \multicolumn{3}{c}{\textbf{Top-1}} 
& \multicolumn{3}{c}{\textbf{Top-5}} 
& \multicolumn{3}{c}{\textbf{Top-10}} \\
\cmidrule(lr){2-4} \cmidrule(lr){5-7} \cmidrule(lr){8-10}
& \textbf{SMR} & \textbf{SGMR} & \textbf{CR}
& \textbf{SMR} & \textbf{SGMR} & \textbf{CR}
& \textbf{SMR} & \textbf{SGMR} & \textbf{CR} \\
\midrule
uFlowCSP ($S{=}5$)       & 68.89 & 64.44 & 58.33 & 71.67 & 71.67 & 65.00 & 74.44 & 73.33 & 66.67 \\
\midrule
Crystal Flow ($S{=}5$)   & 46.67 & 30.00 & 28.89 & 49.44 & 32.22 & 31.11 & 52.78 & 35.00 & 33.89 \\
Crystal Flow ($S{=}100$) & 75.00 & 64.44 & 58.89 & 77.78 & 72.78 & 67.78 & 80.00 & 76.67 & 72.22 \\
\bottomrule
\end{tabular}
\end{table}

\paragraph{Comparison across method families.}
Figure~\ref{fig:tcsp-comparison} places uFlowCSP against both template-based and generative
baselines on the same 180 formulas. Among generative models, five-step uFlowCSP is the
strongest at equal cost and sits close behind CrystalFlow at its full $S=100$ budget, as
quantified above; it is also ahead of EquiCSP. The template method TCSP~2.0
remains the strongest overall, but this comparison is not matched on data access: TCSP~2.0
retrieves from a library two to three orders of magnitude larger than the MP-20 set our
model is trained on, and its accuracy is bounded by that library rather than by a learned
model. The two families are complementary: retrieval wins wherever a close
template already exists, while uFlowCSP reaches competitive symmetry accuracy from the
composition alone, at a fraction of both the data and the inference budget.

\paragraph{Qualitative examples.}
Beyond the overall success rates, Figure~\ref{fig:complex-consensus} shows three
individual cases -- U$_3$Sb$_4$Ir$_3$, Ba$_2$YIrO$_6$, and CeAl$_2$BRu$_2$, where
the prediction matches the ground truth on both structure and space group, drawn in
the same frame for comparison.

\paragraph{Accuracy versus inference cost on MP-20.}
We now turn to the standard MP-20 benchmark, where many published baselines let us
place uFlowCSP on the accuracy--cost frontier. Here uFlowCSP is evaluated under Protocol~B
(see \ref{protocolB}): raw candidates, scored directly with no relaxation or
energy-based selection. This differs from the relaxed, energy-ranked top-5 protocol
of Figure~\ref{fig:ablation-waterfall} and Table~\ref{tab:models}.

uFlowCSP is both accurate and cheap. With five steps at $k=20$ it reaches an
$83.64\%$ match rate, higher than CrystalFlow ($78.34\%$) and DiffCSP ($77.93\%$),
and it does so at a fraction of their cost: $100$ network evaluations per target,
against $2{,}000$ for CrystalFlow and about $20{,}000$ for DiffCSP -- $20$ to
$200\times$ fewer. Even a single step is competitive, matching CrystalFlow at
$78.38\%$ with just $20$ evaluations.

The gain comes from how the budget is spent. Given $100$ evaluations, uFlowCSP
generates several quick candidates and reaches $83.64\%$, whereas CrystalFlow spends
the same budget integrating one candidate finely and reaches $62.02\%$. Adding steps
helps: going from one to five raises the match rate from $78.38\%$ to $83.64\%$ and
lowers RMSE from $0.0783$ to $0.0619$, on par with FlowMM ($0.0566$) and CrystalFlow
($0.0577$), though DiffCSP stays lower at $0.0492$.

The speed follows directly. Generating $10{,}000$ structures takes $0.39$ minutes at
one step and $1.31$ minutes at five, against $6.5$ minutes for CrystalFlow and
$76.1$ minutes for DiffCSP. In short, uFlowCSP matches or beats these baselines on
accuracy while running one to two orders of magnitude faster.
\begin{table}[htbp]
\centering
\caption{Crystal structure prediction on the \textbf{MP-20} benchmark.
$k$ = candidates per target, $S$ = integration
steps, $\mathrm{NFE}=S\times k$ = network evaluations per target, $t$ = wall-time to
generate 10k structures. uFlowCSP scores each target
against only the $k$ candidates generated for it ($n=9046$).
Baseline numbers are as published. Baseline $t$ measured on NVIDIA A800~\citep{luo2025crystalflow}, ours on NVIDIA
A100 (same compute class), batch $=k=20$; $t$ depends on $S$ only. Best per column in \textbf{bold}.
At $k=20$, uFlowCSP reaches 83.64\%, above CrystalFlow at
$k=100$ (82.49) with 100$\times$ fewer network evaluations.}
\label{tab:mp20-main}
\begin{tabular}{lccrccc}
\toprule
Method & $k$ & $S$ & NFE & MR (\%)\,$\uparrow$ & RMSE\,$\downarrow$ & $t$ (min/10k)\,$\downarrow$ \\
\midrule
CDVAE~\citep{xie2022cdvae}             & 1 & $\sim$5{,}000 & $\sim$5{,}000            & 33.90 & 0.1045 & -- \\
DiffCSP~\citep{jiao2023diffcsp}        & 1 & 1{,}000       & \phantom{$\sim$}1{,}000  & 51.49 & 0.0631 & 76.1 \\
FlowMM~\citep{miller2024flowmm}        & 1 & 50--100       & $\sim$100                & 61.39 & 0.0566 & 13.2 \\
CrystalFlow~\citep{luo2025crystalflow} & 1 & 100           & \phantom{$\sim$}100      & 62.02 & 0.0710 & 6.5 \\
\midrule
DiffCSP~\citep{jiao2023diffcsp}        & 20 & 1{,}000 & \phantom{$\sim$}20{,}000 & 77.93          & \textbf{0.0492} & 76.1 \\
CrystalFlow~\citep{luo2025crystalflow} & 20 & 100     & \phantom{$\sim$}2{,}000  & 78.34          & 0.0577          & 6.5 \\
\textbf{uFlowCSP (ours)}                 & 20 & 1       & \phantom{$\sim$}20       & 78.38          & 0.0783          & \textbf{0.39} \\
\textbf{uFlowCSP (ours)}                 & 20 & 5       & \phantom{$\sim$}100      & \textbf{83.64} & 0.0619          & 1.31 \\
\midrule
CrystalFlow~\citep{luo2025crystalflow} & 100 & 100 & \phantom{$\sim$}10{,}000 & 82.49 & 0.0513 & 6.5 \\
\bottomrule
\end{tabular}
\end{table}

\paragraph{Relaxation behavior under a machine-learned potential.}
The match-rate and RMSE results of MP-20 benchmark above summarize how often the raw output is correct,
but not how far a near-miss sits from a stable configuration. To probe that, we examine
how much local relaxation the generated structures require, using
SiO$_2$ as a case study following CrystalFlow~\citep{luo2025crystalflow}. This analysis is a \emph{machine-learned-potential proxy}, not a
first-principles measurement: all structures are relaxed with the ORB-v3 universal
interatomic potential~\citep{rhodes2025orbv3}, and every energy, ionic-step count,
and convergence rate reported in Figure~\ref{fig:orb-relax} is defined by that
potential. The numbers are therefore
internally consistent across the three generators, which are relaxed under an
identical ORB-v3 protocol, but are not directly comparable to the DFT/VASP
quantities reported in the original CrystalFlow SiO$_2$ study, and they inherit any
systematic bias of the underlying potential. We use this diagnostic only as a
cheap, relative measure of how close each generator's raw output sits to a relaxed
local minimum, and not as a physical stability claim. Figure~\ref{fig:orb-relax} shows a clear and consistent trend. The
per-structure energy drop on relaxation (panel a) has a median of $2.41$\,eV/atom
for one-step uFlowCSP ($S{=}1$), which falls to $0.34$\,eV/atom at five steps
($S{=}5$); the five-step generator is thus comparable to the CrystalFlow $S{=}100$
baseline ($0.51$\,eV/atom) on this proxy while using $20\times$ fewer network
evaluations. The relaxation trajectories (panel b) tell the same story: $S{=}1$
structures need the most ionic steps and converge least often under ORB-v3, whereas
$S{=}5$ and CrystalFlow $S{=}100$ converge faster and more completely. After
relaxation, all three generators collapse onto the same low-energy peak (panel c),
indicating that the samplers differ mainly in how far their \emph{raw} outputs sit
from the relaxed basin rather than in the basin they ultimately reach. This is
consistent with the accuracy-cost picture above: additional uFlowCSP steps
sharply improve raw-structure quality, and five steps already yield structures
that relax about as easily as those from a much longer CrystalFlow integration.

\section{Conclusion}

We introduced uFlowCSP, a crystal structure prediction model built on the
one-step MeanFlow formulation. By learning the average instead of the instantaneous
velocity of the probability-flow trajectory, the model generates a complete
periodic structure in a handful of network evaluations, removing the sequential
integration bottleneck that limits diffusion and flow-matching samplers. The
average-velocity field is parameterized by a chemistry- and symmetry-aware
Transformer conditioned on a canonical atom ordering, a global composition
embedding, per-token chemistry descriptors, and a coarse crystal-system
auxiliary token that improves symmetry recovery while being replaced by a null
token at inference, so generation remains formula-only. On the MP-20 benchmark
(Protocol~B), uFlowCSP reaches a 83.64\% match rate at
$k=20$ with only 100 network evaluations per target, surpassing CrystalFlow
(78.34\% at 2{,}000 evaluations) and DiffCSP (77.93\% at $\sim$20{,}000) while
cutting wall-clock generation time by 5 to 58 times. On the 180-formula
CSPBench suite (Protocol~A), the model attains 72/72/65\% structure,
space-group, and consensus match rates at only five sampling steps; a CrystalFlow
baseline held to the same budget reaches just 49/32/31\%, and exceeds our model
only when its integration budget is raised twenty-fold, showing that the
average-velocity formulation gains accuracy per network evaluation rather than
at peak accuracy. Ablations show that the four conditioning signals
contribute additively to model performance.
Together these results place CSP at a new point on the accuracy-cost frontier,
demonstrating that high-quality crystal structures can be generated in a single
or few network evaluations without sacrificing structural or symmetry fidelity.

Finally, our best-performing configuration adopts a deliberately simple
geometric representation, and we report a negative result on the more
crystallographically natural alternative. uFlowCSP transports
Cartesian positions and a raw nine-degree-of-freedom $3\times 3$ lattice along
Euclidean paths from a Gaussian prior, ignoring both the periodicity of
fractional coordinates and the over-completeness of the unreduced cell. We therefore
present this as a limitation of our current implementation rather than a general
claim about periodic generative geometry; isolating the two components, in
particular pairing torus coordinates with an orientation-free but continuous
lattice representation, is the most direct next step.

\section{Contribution}
Conceptualization, J.H.; methodology, J.H., S.D,  L.W., S.O., D.D.; software, S.D; resources, J.H.; writing, original draft preparation, J.H., S.D; writing and review and editing, J.H., S.D, D.D, L.W, S.O.; visualization, S.D.; supervision, J.H.;  funding acquisition, J.H.

\section*{Acknowledgement}
The research reported in this work was supported in part by National Science Foundation under the grant and 2110033, 2311202, and 2320292. The views, perspectives, and content do not necessarily represent the official views of the NSF.

\bibliographystyle{iclr2026_conference}  
\bibliography{references}

\end{document}